\documentclass[12pt]{article}

\newcommand{\ra}{\rightarrow} 
\newcommand{\nn}{\nonumber}

\newcommand{\al}{\alpha}

\newcommand{\ov}{\overline} 
\newcommand{\oz}{{\ov{z}}}
 
\newcommand{\diag}{{\rm diag}}

\title{\vspace{-1in}\parbox{\linewidth}{\small\hfill\shortstack{IASSNS-HEP-98/34}}
\vspace{0.6in}\\
Solution of N=2 Gauge Theories via Compactification to Three Dimensions}
\author{
Anton Kapustin\thanks{Research supported in part by DOE grant DE-FG02-90-ER40542}\\
{\sl\small School of Natural Sciences, Institute for Advanced Study}\\
{\sl\small Olden Lane, Princeton, NJ 08540}}

\begin{document}
\begin{titlepage}
\renewcommand{\thepage}{ }
\renewcommand{\today}{ }
\thispagestyle{empty}

\maketitle

\begin{abstract}
A number of $N=2$ gauge theories can be realized by brane configurations in
Type IIA string theory. One way of solving them involves
lifting the brane configuration to M-theory. In this paper we present an alternative 
way of analyzing a subclass of these theories (elliptic models). We observe that upon 
compactification on a circle one can use a version
of mirror symmetry to map the original brane configuration into one containing
only D-branes. Simultaneously the Coulomb branch of the four-dimensional
theory is mapped to the Higgs branch of a five-dimensional theory with three-dimensional
impurities. The latter does not receive quantum corrections and can
be analyzed exactly. The solution is naturally formulated in terms of an
integrable system, which is a version of a Hitchin system on a punctured torus. 
 
\end{abstract}
\end{titlepage}

\section{Introduction}
In the last few years a lot of effort has been invested into studying moduli spaces
of vacua of supersymmetric gauge theories.
This problem is quite nontrivial when the theory in question is strongly coupled in the 
infrared. It is remarkable
that nevertheless a complete solution has been found for a large number of $N=2$ gauge
theories using various techniques. In particular, in Ref.~\cite{Witten} the
structure of the Coulomb branch of theories with product gauge
groups $SU(n_1)\times SU(n_2)\times \ldots \times SU(n_{k-1})$ has been analyzed.
The idea is to realize the gauge theory 
as a low-energy theory on a configuration of branes in IIA string theory.
To this end one considers an array of $k$ parallel NS5-branes extending in the 
$x^0,x^1,x^2,x^3,x^4,x^5$ directions and having coincident positions in the $x^7,x^8,x^9$
directions. Then one suspends $n_\al$ parallel D4-branes between the $\al^{\rm th}$ and 
$\al+1^{\rm st}$ 5-branes, so that the world-volume of D4-branes is extended in the 
$x^0,x^1,x^2,x^3,x^6$ directions. (One can also include D6-branes, but we do not consider
this possibility here). Since the extent of this configuration in the $x^6$
direction is finite, the low-energy theory is a four-dimensional gauge
theory living in the $x^0,x^1,x^2,x^3$ plane. It can be checked that this arrangement of
branes leaves eight supersymmetries unbroken, so we get at least
$N=2$ supersymmetry in $d=4$. Naively, the gauge group appears
to be $U(n_1)\times U(n_2)\times\cdots\times U(n_{k-1})$ with matter hypermultiplets in the
bifundamental representation (i.e. the hypermultiplets transform as
$(n_1,\ov{n}_2)\oplus (n_2,\ov{n}_3)\oplus\cdots\oplus (n_{k-2},\ov{n}_{k-1})$.) The
hypermultiplets come from open strings connecting neighboring stacks of D4-branes.
A more detailed analysis~\cite{Witten} shows that the center-of-mass motion of these 
stacks is ``frozen out'', in the sense that the $x^4,x^5$ positions of the centers-of-mass
are parameters of the Lagrangian rather than dynamical fields. As a
consequence the gauge group is actually 
$SU(n_1)\times SU(n_2)\times \cdots \times SU(n_{k-1})$. It is convenient to combine the
$x^4,x^5$ coordinates of the center-of-mass of the $\al^{\rm th}$ stack into a complex 
number $v_\al=x^4_\al+ix^5_\al$. Then the bare mass of the $\al^{\rm th}$ hypermultiplet 
is given by $v_{\al+1}-v_\al$.

In order to solve for the Coulomb branch of this theory, one lifts the brane configuration
to M-theory. If the IIA string theory is strongly coupled, the radius of the
M-theory circle (whose coordinate we will call $x^{10}$) is large. In this case the IIA brane
configuration lifts to
a single large smooth 5-brane with worldvolume ${\bf R}^{1,3}\times C$.
Here ${\bf R}^{1,3}$ has coordinates $x^0,x^1,x^2,x^3$, and $C$ is
a Riemann surface holomorphically embedded in ${\bf C}\times {\bf C}^*$ with coordinates
$v=x^4+ix^5,t=\exp(x^6+ix^{10})$. The moduli of the 
field theory translate into the
moduli of the embedding $C\ra {\bf C}\times {\bf C}^* $. Of course, the embedding is not
arbitrary: one must require that upon forgetting $x^{10}$ the M-theory 5-brane project back
to the correct IIA configuration. To solve the model one 
has to find a holomorphic family of embeddings satisfying this constraint and depending on 
the right number of parameters.
 
Another way of solving $N=2$ gauge theories is based on their relation to complex integrable
system~\cite{GKMMM,MW,DW}.  We will explain the precise meaning of this statement in the next
section.  For now it suffices to say that a complex integrable system can be thought of as a
complexification of an integrable system of classical mechanics, and that the solution of
every $N=2$ gauge theory is encoded in some complex integrable system.  Therefore one can try
to match the known integrable models with particular $N=2$ gauge theories.  The first success
of this approach was the observation of Ref.~\cite{MW} that an affine Toda chain associated
with a simple group $G$ provides a solution for the $N=2$ gauge theory with gauge group
$\hat{G}$ (the Langlands dual of $G$) and no hypermultiplets.  Soon after that the $SU(n)$
Hitchin system~\cite{Hitchin} on a torus with a single puncture was shown to give a solution
of the $SU(n)$ gauge theory with a massive adjoint hypermultiplet~\cite{DW}.  Some other
matches were suggested in Ref.~\cite{Russians}.

It is natural to ask about the relation between the two approaches. An answer suggested in
Ref.~\cite{Witten} is that a family of embeddings of a Riemann surface $C$ into 
${\bf C}\times {\bf C}^*$ can be thought of as an integrable system too, namely the 
integrable system of Donagi-Markman type~\cite{DM}. Thus every model solved in 
Ref.~\cite{Witten} is matched to an integrable system.

We do not consider this a completely satisfactory answer, however. Indeed, in the cases 
where both approaches are applicable, the solutions seem totally different, and some effort
is required to see that they are equivalent. For example, the fact that the
solution of the $SU(n)$ theory with a massive adjoint hypermultiplet can be represented
by a Hitchin-type system looks like a miracle from the point of view of M-theory.

In this paper we would like to address this problem by suggesting a new method of solving
$N=2$ gauge theories.\footnote{Similar ideas were considered in Ref.~\cite{Gukov}.}  The
starting point is again the brane configuration of Ref.~\cite{Witten}, but with $x^3$
compactified on a circle.  Thus we will be studying $N=2$ gauge theories on ${\bf
R}^{1,2}\times S^1$.  For technical reasons we will consider only elliptic models, i.e.  those
with $x^6$ direction compactified as well.  However, it will be clear from the derivation that
the method can be applied more generally.  Thus the gauge theories we are solving are finite
and have gauge group $SU(n)_1\times SU(n)_2\times\cdots\times SU(n)_k\times U(1)$ and matter
in the bifundamental.  The gauge group and the matter content of this theory can be encoded in
an affine $A_{k-1}$ quiver.  It turns out there is a version of mirror symmetry (in the sense
of Ref.~\cite{IS}) which maps the Coulomb branch of the original (``electric'') theory to the
Higgs branch of a certain five-dimensional ``magnetic'' theory on ${\bf R}^{1,2}\times T^2$,
with three-dimensional impurities localized at points on $T^2$.  The Higgs branch does not
receive quantum corrections and can be analyzed exactly.  This yields a solution of the
original problem, for any compactification radius.  An interesting feature of this approach is
that the solution is automatically encoded in a Hitchin-type integrable system.  Thus, at
least for the class of models considered in this paper (elliptic models), the corresponding
Donagi-Markman systems arise from Hitchin-type systems.  In particular, we rederive the fact
noticed in Ref.~\cite{DW} that the $SU(n)$ gauge theory with a massive adjoint is solved by an
$SU(n)$ Hitchin system on a punctured torus.  Another spin-off of our approach is a new
explanation of the ``freezing'' of $U(1)$ factors in the d=4 gauge theories noticed in
Ref.~\cite{Witten}.

In the next section we give a very brief summary of the relation between $N=2$ gauge 
theories and integrable systems. In this we follow Ref.~\cite{DW}. In section 3 we
discuss the mirror transform alluded to above. The analysis of the Higgs branch of 
the ``magnetic'' theory is presented in section 4. There we also compare our solution
with that obtained in Ref.~\cite{Witten}. Some idiosyncratic remarks are collected in
section 5.

\section{N=2 Gauge Theories and Integrable Models} 
In this section we remind the reader the
relation between low-energy Lagrangians of $N=2$ gauge theories and integrable models of
classical mechanics~\cite{DW}.  Consider an $N=2$ gauge theory with gauge group $G$ of rank
$r$ and matter hypermultiplets.  The Coulomb branch of the theory $U$ is a special K\"ahler
manifold of complex dimension $r$.  As a complex manifold, $U$ is a copy of ${\bf C}^r$.  The
special K\"ahler metric is encoded in a holomorphic function ${\cal F}$ on $X$, the
prepotential.  This function is multi-valued, however, so it is desirable to give a
description of the metric in more invariant terms.  To this end one considers a fibration
$\pi:X\ra U$, where $X$ is a complex manifold of dimension $2r$ and the fibers of $\pi$ are
Abelian varieties $A_r$ of dimension $r$.  (In other words, $A_r$ is a complex torus together
with a $(1,1)$-form $t$ (polarization) which is positive and has integral periods).  We will
call the fibration $\pi:X\ra U$ the Seiberg-Witten fibration.  One also needs a closed
holomorphic $(2,0)$-form $\omega$ on $X$ whose restriction to the fibers of $\pi$ is zero.
Together these data define a metric on $U$ in the following manner:  one takes the
$(r+1,r+1)$-form $t^{r-1}\wedge\omega\wedge\ov{\omega}$ and integrates over the fibers of
$\pi$; this yields a $(1,1)$-form on $U$ which is the K\"ahler form of the metric we are
after.  An additional physical requirement is that this metric be nondegenerate away from
singular fibers of $\pi$.  This is achieved by asking that $\omega$ be nondegenerate away 
from the singular fibers.

So far we described how the fibration $\pi:X\ra U$ together with $\omega$ define a K\"ahler 
metric on $U$ and therefore the low-energy effective action for massless scalars. The 
low-energy theory of the Coulomb branch also contains $r$ photons. 
To define their action one needs to specify the ``$\tau$-parameter'',
i.e. an $r\times r$ matrix which is a holomorphic function on $U$ whose imaginary part is
positive-definite (away from singular fibers). The
``$\tau$-parameter'' encodes the gauge couplings and theta-angles of the photons, and
because of electric-magnetic duality it is defined up to $Sp(2r,{\bf Z})$ transformations.
Given the Seiberg-Witten fibration it is very easy to read off $\tau$: at a point $u\in U$
it is the complex structure of the fiber $\pi^{-1}(u)$.

One can think of $X$ together with $\omega$ as a complex symplectic manifold, i.e.
as a complexification of the phase space of some mechanical system with $r$ degrees
of freedom. Moreover, since the restriction of $\omega$ to the fibers of $\pi$ is
zero, any two functions on $U$ Poisson-commute. Therefore the coordinates on $U$
form a maximal set of commuting integrals of motion, and the corresponding mechanical
system is integrable (it fulfills the conditions of Liouville's theorem). In other words,
the coordinates on $U$ are action variables, while the coordinates on the fiber $A_r$
are the canonically conjugate angle variables. Thus any $N=2$ gauge theory
corresponds to a certain complex integrable system of classical mechanics.

\section{Compactification to Three Dimensions and the Mirror Transform}

We start with the brane configuration considered in Ref.~\cite{Witten}: $k$ NS5-branes
located at $x^7=x^8=x^9=0$ and at $x^6=s_1,\ldots,s_k$ and $n$ D4-branes located at
$x^7=x^8=x^9=0$ and worldvolume parametrized by $x^0,x^1,x^2,x^3,x^6$. The $x^6$ coordinate
is taken to be periodic with period $2\pi L$. Thus D4-branes are wrapped on a circle of
radius $L$. Since D4-branes can end on NS5-branes, a stack of $n$ parallel D4-branes
can split at NS5-branes, producing $k$ independent stacks of D4-branes. They can move
in the $(x^4,x^5)$ plane, which we regard as a copy of ${\bf C}$ parametrized by $v=x^4+ix^5$.
This is the Coulomb
branch of the theory. (There is also a mixed Higgs-Coulomb branch corresponding to the
situation when some or all D4-branes move off the $x^7=x^8=x^9=0$ plane, but we will
not consider it here.) Let us denote the center-of-mass coordinate of the $\al^{\rm th}$ stack by
$v_\al$. It was shown in Ref.~\cite{Witten} that moving the center-of-mass of
any stack in the $x^4,x^5$ directions relative to other stacks costs an infinite amount of energy,
therefore the relative center-of-mass coordinates $m_\al=v_{\al+1}-v_\al,\ \al=1,\ldots,k$ 
are parameters of the theory rather than moduli. Consequently the low-energy
theory on D4-branes is a $d=4, N=2$ theory with gauge group $SU(n)_1\times\cdots\times 
SU(n)_k\times U(1).$
There are also $k$ bifundamental hypermultiplets $Q_\al,\al=1,\ldots,k$; $Q_\al$ transforms
as $n$ with respect to $SU(n)_\al$ and as $\ov{n}$ with respect to $SU(n)_{\al+1}$.
They are not charged with respect to $U(1)$. These hypermultiplets arise from open strings
connecting the adjacent stacks of D4-branes, therefore the bare mass of $Q_\al$ is $m_\al.$

We now compactify $x^3$ on a circle of radius $R$. The theory becomes effectively
$2+1$-dimensional at energies lower than $1/R$. Its Coulomb branch is a
hyperk\"ahler manifold $X$ of dimension $4r=4(kn-k+1)$ with a distinguished complex 
structure in which it looks like a fibration $\pi: X\ra {\bf C}^r$ with fibers being
abelian varieties $A_r$ of complex dimension $r$~\cite{SW}. Let us remind
how this comes about. The picture is most clear when $R$ is large compared to all field 
theory scales. One can then go to a low-energy limit in $d=4$ and then compactify to
$d=3$. The $r$ photons of the Coulomb branch in $d=4$ reduce to $r$ photons plus
$r$ scalars in $d=3$. The scalars live in a torus whose size scales as $1/\sqrt{R}$,
as they originate from Wilson lines around the compact direction. Furthemore, $r$ photons
in $d=3$ are dual to $r$ compact scalars. Thus we get $r$ more scalars also living on 
a torus of size of order $1/\sqrt{R}$. If one uses the distinguished complex structure,
these $2r$ real scalars can be thought of as $r$ complex scalars taking values in a 
complex torus $A_r$. One can show that in this distingushed complex structure the total 
space of the fibration $X\ra {\bf C}^r$ is the same as the Seiberg-Witten fibration of
the parent theory in $d=4$. This result is most easily derived for large $R$; it is then
true for any $R$, because the complex structure of $X$ does not depend
on $R$~\cite{SW}. 

Since $X$ is hyperk\"ahler, it also carries a complex symplectic $(2,0)$-form 
$\omega_2+i\omega_3$ and a $(1,1)$ K\"ahler form $\omega_1$. For large $R$ the periods of
$\omega_1$ evaluated on the 2-cycles of $A_r$ are proportional to $1/R$, since the 
linear size of $A_r$ scales as $1/\sqrt{R}$. Furthemore, the form $\omega_2+i\omega_3$ 
coincides with the form $\omega$ which was a part of the four-dimensional data~\cite{SW} 
(see section 2). 

We now wish to identify the Coulomb branch of our $d=3$ theory with the Higgs branch of 
some ``magnetic'' theory. To this end we perform T-duality on $x^3$, then IIB S-duality,
and then again T-duality on $x^3$. As a result the string coupling $\lambda$ is mapped 
to $\tilde{\lambda}=R^{3/2}\lambda^{-1/2}$, and $R\ra \tilde{R}=(R\lambda)^{1/2},\ L\ra 
\tilde{L}=L(R/\lambda)^{1/2}$.\footnote{We set $\alpha'=1$.}
Notice also that after the dualities the modular parameter of the torus in the $x^3,x^6$
directions becomes $\tilde{\tau}\equiv\tilde{L}/\tilde{R}=L/\lambda.$  

After the first T-duality NS5-branes become IIB NS5-branes. S-duality turns them
into IIB D5-branes, and the second T-duality turns them into IIA D4-branes located at 
fixed $x^3,x^6,x^7,x^8,x^9$. We will refer to them as D4$'$-branes. To understand what 
happens with D4-branes, we consider
the situation when D4-branes are not broken at the NS5-branes. This corresponds to the 
origin of the classical moduli space in the original (``electric'') theory, where the 
Coulomb and Higgs branches meet.
In this case it is easy to see that D4-branes are left unchanged by this sequence of 
dualities, i.e. they remain D4-branes wrapped around the $T^2$ parametrized by $x^3,x^6$. 
The D4$'$-branes are localized at points of this $T^2$. Thus 
the ``magnetic'' theory is a $d=5$ theory on ${\bf R}^{1,2}\times T^2$ with impurities 
localized
at points on $T^2$. At energies much lower than ${\rm min}(1/\tilde{R},1/\tilde{L})$ it 
becomes a $d=3$ theory. The gauge group of the ``magnetic'' theory is $U(n)$, and, if not 
for the impurities, it would have sixteen unbroken supercharges. The impurites break half 
of supersymmetries and give rise to $k$ fundamental hypermultiplets localized at points 
on $T^2$. The hypermultiplets come from open strings connecting D4 and D4$'$-branes. 

Now we can understand how the flat directions in the ``electric'' and ``magnetic'' 
theories are matched.
In the ``electric'' picture the Coulomb branch was characterized by the fact that 
D4-branes could not move off in the $x^7,x^8,x^9$ directions. In the ``magnetic'' 
picture this occurs when fundamental hypermultiplets have VEVs, higgsing the gauge group.
Thus the mirror
transform maps the Coulomb branch of the ``electric'' theory to the Higgs branch of the 
``magnetic'' theory. Conversely, the ``electric'' Higgs branch corresponds to the
``magnetic'' Coulomb branch.

Here is another way to describe this ``mirror transform.'' We reinterpret the  
``electric'' IIA brane 
configuration as an arrangement of branes in M-theory compactified on $T^3$ with 
coordinates $x^3,x^6,x^{10}$. D4-branes are interpreted as M5-branes wrapped around $T^3$, 
while NS5-branes lift to M5-branes wrapped around $x^3$ and localized in $x^6,x^{10}.$ 
The mirror transform amounts to reinterpreting $x^3$ as the M-theory circle. 
The ``magnetic'' IIA picture involves $n$ D4-branes wrapped around a $T^2$ with coordinates
$x^6,x^{10}$, and $k$ D4$'$-branes localized
at points of this $T^2$ (see Figure 1).
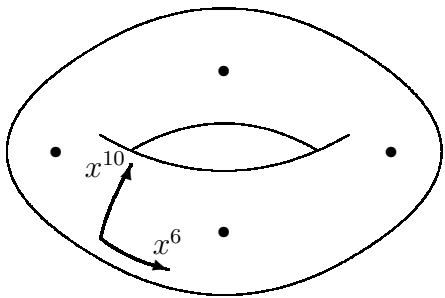
\begin{figure}

\setlength{\unitlength}{0.2em}
\begin{center}
\begin{picture}(100,40)

\qbezier(30,2.68)(50,-8.87)(70,2.68)
\qbezier(30,37.32)(50,48.87)(70,37.32)
\qbezier(30,2.68)(0,20)(30,37.32)
\qbezier(70,2.68)(100,20)(70,37.32)
\qbezier(30,22.68)(50,11.13)(70,22.68)
\qbezier(35,20.21)(50,28.87)(65,20.21)
\thicklines
\qbezier(30,6)(31,11)(35,18)
\qbezier(30,6)(35,2.3)(41,1)
\put(35.1,18.1){\vector(1,3){0}}
\put(41.2,0.9){\vector(3,-1){0}}
\put(31,18){\makebox(0,0){$x^{10}$}}
\put(41,3.5){\makebox(0,0)[b]{$x^6$}}

\put(50,7){\circle*{1.5}}
\put(50,33){\circle*{1.5}}
\put(23,20){\circle*{1.5}}
\put(77,20){\circle*{1.5}}

\end{picture}\end{center}
\caption{After $x^3$ is reinterpreted as the M-theory circle, the ``electric'' D4-branes
become D4-branes wrapped around $x^6$ and $x^{10}$, while NS5-branes become D4$'$-branes
localized in the $x^6,x^{10}$ directions. Here the positions of the D4$'$-branes are
shown as punctures on the $T^2$ parametrized by $x^6,x^{10}$.}

\end{figure}
This way of thinking about the mirror transform 
allows one to see some facts about the ``magnetic'' theory more easily.  
For example, consider the ``magnetic'' Wilson line on the worldvolume of D4-branes 
around a puncture created by a D4$'$-brane. What is its analogue in the ``electric'' theory?
The Wilson line around the puncture can be alternatively computed as the holonomy
around the composite contour shown in Figure 2. 
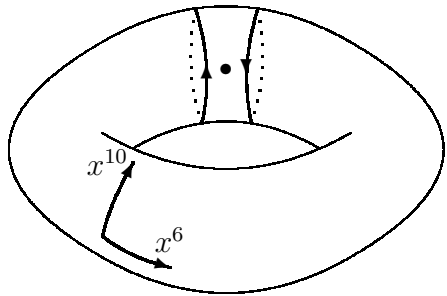
\begin{figure}

\setlength{\unitlength}{0.2em}
\begin{center}
\begin{picture}(100,40)

\qbezier(30,2.68)(50,-8.87)(70,2.68)
\qbezier(30,37.32)(50,48.87)(70,37.32)
\qbezier(30,2.68)(0,20)(30,37.32)
\qbezier(70,2.68)(100,20)(70,37.32)
\qbezier(30,22.68)(50,11.13)(70,22.68)
\qbezier(35,20.21)(50,28.87)(65,20.21)
\thicklines
\qbezier(30,6)(31,11)(35,18)
\qbezier(30,6)(35,2.3)(41,1)
\put(35.1,18.1){\vector(1,2){0}}
\put(41.2,0.9){\vector(3,-1){0}}
\put(31,18){\makebox(0,0){$x^{10}$}}
\put(41,3.5){\makebox(0,0)[b]{$x^6$}}

\put(50,33){\circle*{1.5}}
\qbezier(45,42.6)(48,33)(46,24.3)
\qbezier(55,42.6)(52,33)(54,24.3)
\put(46.9,34){\vector(0,1){0}}
\put(53.3,32){\vector(0,-1){0}}
\qbezier[10](45,42.6)(43,33)(46,24.3)
\qbezier[10](55,42.6)(57,33)(54,24.3)

\end{picture}\end{center}
\caption{The Wilson line around the puncture created by the D4$'$-brane can be 
deformed to a product of the Wilson lines along the two contours shown in the figure.}

\end{figure}
In the M-theory language, the
Wilson line on the D4-brane along each component of this contour is interpreted as a 
flux of B-field through an appropriate two-cycle. In the case of the contours shown in 
Figure 2 each two-cycle is a $T^2$ parametrized by $x^3,x^{10}$. Now to go back to the
``electric'' picture we reinterpret $x^{10}$ as the M-theory circle. Then
each of the two-cycles becomes a contour going around $x^3$, one of them to the left
of the NS5-brane, another one to the right of it. Thus the ``magnetic'' Wilson line
around the puncture is mapped to the jump of the ``electric'' Wilson lines
around $x^3$ as one traverses the NS5-brane. Actually,
in an ordinary compactification from $d=4$ to $d=3$ we do not allow for such jumps
of the ``electric'' Wilson line. This corresponds to the ``magnetic'' Wilson line around 
the puncture being trivial. This will play an important role in the discussion of 
the next section. One $can$ consider more general compactifications where the ``electric''
Wilson line jumps at the locations of the NS5-branes. These jumps are then the additional 
parameters of the compactified theory. However, since we are mainly interested
the $d=4$ theory, we will only consider the usual compactification with no jumps.

The advantage of the ``magnetic'' description is obvious: the Higgs branch cannot receive 
string loop corrections~\cite{IS}, and therefore can be computed classically. A question 
may arise how a classical computation on the ``magnetic'' side manages to capture quantum 
effects on the ``electric'' side. The reason is that the classical answer depends 
nontrivially on $\tilde{\tau}$, the modular parameter of $T^2$. 
We have seen that it is given by $\tilde{\tau}=L/\lambda$, which coincides
with the effective four-dimensional gauge coupling of the ``electric'' theory. The metric on 
the Higgs branch will remember about $\tilde{\tau}$ as long as we treat the ``magnetic'' 
theory as a five-dimensional theory with three-dimensional impurities, and do not reduce it 
to three dimensions.

In the limit $R\ra 0$, when the ``electric'' theory becomes truly three-dimensional, our 
mirror transform reduces to that discussed in Refs.~\cite{Berkeley,HW}. Indeed, suppose we 
take $R\ra 0$ keeping the $d=3$ gauge coupling $LR/\lambda$ fixed. The ``electric'' theory
reduces in this limit to a $d=3,N=4$ theory with the same gauge group and matter
content as in $d=4$. To see what happens on the ``magnetic'' side, notice
that $\tilde{R}/\tilde{L}=\lambda/L\ra 0$. This means that the $T^2$ of the impurity theory
becomes very ``thin'' in the $x^3$ direction. Then it is helpful to perform T-duality
on $x^3$. The resulting brane configuration consists of $n$ D3-branes parallel to
$x^0,x^1,x^2,x^6$ and $k$ D5-branes parallel to $x^0,x^1,x^2,x^3,x^4,x^5$. According
to Refs.~\cite{Berkeley,HW} this setup produces the right ``magnetic'' theory.

\section{Solution of the Models}

\subsection{The Higgs Branch of the Impurity Theory}

In this section we analyze the Higgs branch of the ``magnetic'' theory. The D-flatness
conditions for an equivalent impurity model have been derived in Ref.~\cite{KS}.
(Our model is related to that considered in section 2.3 of Ref.~\cite{KS} by T-duality
on $x^1,x^2$.) The result is that the Higgs branch is given by the moduli space of
solutions of the following partial differential equations on $T^2$:
\begin{eqnarray}\label{master}
&&F_{z\ov{z}}-[\Phi_z,\Phi_z^\dagger]=
\frac{\pi}{RL} \sum_{\al=1}^k \delta^2 (z-z_\al)\ (Q_\al\otimes Q_\al^\dagger -
\tilde{Q}_\al^\dagger\otimes\tilde{Q}_\al), \nn \\
&&\ov{D}\Phi_z\equiv \ov{\partial}\Phi_z-[A_\oz,\Phi_z]=-\frac{\pi}{RL}\sum_{\al=1}^k 
\delta^2 (z-z_\al)\ Q_\al\otimes\tilde{Q}_\al.
\end{eqnarray}
Here $A_z dz + A_\oz d\oz$ is a $U(n)$ connection on $T^2$, $F_{z\ov{z}}=
\partial A_\oz-\ov{\partial} A_z - [A_z,A_\oz]$, $\Phi_z$ is a complex adjoint-valued 1-form,
and $Q_\al,\tilde{Q}_\al,\al=1,\ldots,k$ are complex variables. It is understood that
$A_z=-A_\oz^\dagger$. 
The objects appearing
in these equations have a transparent meaning: $A$ is part of the gauge field living
on the worldvolume of D4-branes, $\Phi_z$ is the adjoint field parametrizing the motion
of the D4-branes in the directions parallel to D4$'$-branes ($x^4$ and $x^5$), while
$(Q_\al,\tilde{Q}_\al)$ make up the fundamental hypermultiplet localized at the 
$\al^{\rm th}$ impurity. In the absence of impurity terms, Eqs.~(\ref{master}) are known as
Hitchin equations. Hitchin equations are self-duality equations reduced to two dimensions.
The effect of the impurity terms is to introduce poles for $\Phi_z$
and $A_z$ at $z=z_\al,\al=1,\ldots,k$. The residues are not fixed; rather they are
determined by the VEVs of the fundamentals.

As usual in $N=2$ theories, D-flatness conditions can be thought of as moment map
equations for a hyperk\"ahler quotient~\cite{KS}. In this case the group ${\cal G}$ is
infinite-dimensional: it is a group of smooth maps $T^2\ra U(n)$. It acts on a space $Y$
consisting of all sets of the form $(A_\oz,\Phi_z,Q_1,\tilde{Q}_1,\ldots,Q_k,\tilde{Q}_k)$.
The space $Y$ is an infinite-dimensional affine space with a natural hyperk\"ahler
structure. An element $g(z)\in{\cal G}$ acts on $Y$ in the following manner:
\begin{eqnarray}
&& A_\oz(z)\ra g(z)(z)A_\oz(z)g^{-1}(z)+g(z)\ov{\partial} g^{-1}(z),\qquad
\Phi_z\ra g(z)\Phi_z(z)g^{-1}(z), \nn\\
&&Q_\al\ra g(z_\al) Q_\al,\qquad \tilde{Q}_\al\ra \tilde{Q}_\al g(z_\al)^{-1},\qquad \al=1,\ldots,k.
\end{eqnarray}
A short computation reveals that the submanifold in $Y$ defined by Eqs.~(\ref{master}) 
is precisely the zero level of the moment map for ${\cal G}$. Thus the moduli space
of Eqs.~(\ref{master}) is the hyperk\"ahler quotient of $Y$ with respect to ${\cal G}$ at
zero level. A formal application of the theorem of Hitchin, Karlhede, Lindstr\"om and
Ro\v{c}ek~\cite{HKLR} tells us that $X$ has a natural hyperk\"ahler metric inherited from
that on $Y$. This is a formal statement, because $Y$ is infinite-dimensional, and the
questions of existence of various objects needed in the construction of Ref.~\cite{HKLR}
are nontrivial. 

As a matter of fact, one immediately sees two problems with our definition of $X$, 
due precisely to the formal character of our manipulations.
First, notice that Eqs.~(\ref{master}) imply that $\Phi_z$ and $A_z$ have simple
poles at $z=z_\al,\al=1,\ldots,k,$ with variable residues. Therefore the variations
of $\Phi_z$ and $A_z$ generically have simple poles as well, and their norm is
logarithmically divergent. As a consequence, some of the  
tangent vectors to $X$ have infinite norm, since they correspond to 
zero modes of Eqs.~(\ref{master}) which are not normalizable on $T^2$. This means
that it costs an infinite amount of energy to move in these directions on $X$.
Second, we saw in the previous
section that the Wilson lines around the punctures at $z=z_\al$ must be trivial. Then $A_z$
must be nonsingular at $z=z_\al$, and the first equation in Eq.~(\ref{master}) implies that
\begin{equation}\label{realc}
Q_\al\otimes Q_\al^\dagger -
\tilde{Q}_\al^\dagger\otimes\tilde{Q}_\al=0, \qquad \al=1,\ldots,k.
\end{equation}
But this restriction is incompatible with the hyperk\"ahler structure of $X$.

It is clear what the resolution should be. The residues of $\Phi_z$ and $A_z$ are parameters
of the theory rather than dynamical fields. The true moduli correspond to those variations 
which have finite norm. To identify the true moduli space one has to freeze the residues of
$A_z$ and $\Phi_z$
at some values, so that variations of $\Phi_z$ and $A_z$ are nonsingular. Then the remaining
tangent vectors will have finite norm, and the metric will be well-defined. In fact, 
the constraint Eq.~(\ref{realc}) does just this, freezing the residue of $A_z$ at zero value. 
This eliminates some of the tangent vectors with infinite norm (the ones corresponding to
variations of $A_z$ alone). Further constraints are necessary to kill the nonnormalizable
tangent vectors associated with variations of $\Phi_z$. To figure out the precise form
of these constraints, let us use the $U(n)$ gauge transformations to bring all $Q$'s
and $\tilde{Q}'s$ to the form
\begin{equation}\label{Qform}
Q_\al=(a_\al,b_\al,0,\ldots,0), \qquad \tilde{Q}_\al=(\tilde{a}_\al,0,0,\ldots,0),
\qquad \al=1,\ldots,k.
\end{equation}
Eq.~(\ref{realc}) implies that $b_\al=0, |a_\al|=|\tilde{a}_\al|, \al=1,\ldots,k.$
According to the second of Eqs.~(\ref{master}), for the residues of $\Phi_z$ to be fixed
one has to impose an additional constraint
\begin{equation}\label{complexc}
a_\al\tilde{a}_\al=m_\al, \qquad \al=1,\ldots,k,
\end{equation}
where $m_\al$ are complex constants. These constraints are still invariant with respect to
$U(n)$ gauge transformations which reduce to $U(1)\times U(n-1)$ at $z=z_\al$. One can use 
the $U(1)$ part of these gauge transformations to make all $a_\al$ real. Then the degrees
of freedom corresponding to $Q$'s and $\tilde{Q}$'s are completely frozen, and we are left
with Hitchin equations for $A_z$ and $\Phi_z$ with fixed residues for $\Phi_z$.

To summarize, the slice of $X$
which does have a well-defined metric consists of solutions of the Hitchin equations
\begin{eqnarray}\label{master'}
&&F_{z\ov{z}}-[\Phi_z,\Phi_z^\dagger]=0, \nn \\
&&\ov{D}\Phi_z=-\frac{\pi}{RL}\sum_{\al=1}^k 
\delta^2 (z-z_\al)\ \diag (m_\al,0,\ldots,0),
\end{eqnarray}
modulo the gauge group ${\cal G}_0$ of $U(n)$ gauge transformations which reduce to $U(n-1)$ 
at $z=z_\al$. Let us call this moduli space $X_0$. The preceeding discussion shows that
$X_0$ has a good metric, unlike $X$. Moreover, Eqs.~(\ref{master'}) have the form of
moment map equations for ${\cal G}_0$, hence the metric on $X_0$ is hyperk\"ahler.
Thus both problems mentioned above have been resolved.

It will be shown in section 4.3 that $\dim X_0=4(kn-k+1)$. This is the right dimension
for the Coulomb branch of the $SU(n)^k\times U(1)$ gauge theory.
Recall that the gauge group is $SU(n)^k\times U(1)$ and not $U(n)^k$ because, as observed
in Ref.~\cite{Witten}, 
it costs an infinite amount of energy to excite the $k-1$ would-be moduli corresponding to
the missing $U(1)$'s.
Consequently, the VEVs of these moduli are actually parameters of the theory,
the bare masses of the bifundamentals. There is a total of $k$ bifundamentals, but
the $k$ mass parameters satisfy a constraint
\begin{equation}
\sum_{\al=1}^k m_\al=0.
\end{equation} 
Where are these parameters in the ``magnetic'' description? Since we cunningly denoted
the parameters in Eq.~(\ref{master'}) with the same symbol $m_\al$, the answer should be
obvious.\footnote{Of course, what we really claim is that such identification holds up to
an overall multiplicative factor. But since the ``electric'' theory is finite, one can set 
this factor to one by a choice of scale.}
As a simple check of this identification note that the trace of the second
of Eqs.~(\ref{master'}) implies that ${\rm Tr}\ \Phi$ is a meromorphic function with simple
poles at $z=z_\al$ and residues $im_\al/2RL$. Hence $\sum_\al m_\al=0$. Another check
is that the $U(1)_R$ symmetry of the $N=2,$ $d=4$ theory is realized geometrically in the
brane configuration as a rotation in the $x^4,x^5$ plane. Consequently, a
$U(1)_R$ transformation acts on $\Phi$ by $\Phi\ra e^{i\phi}\,\Phi$, and therefore by virtue
of Eqs.~(\ref{master'}) $m_\al\ra e^{i\phi}\, m_\al$.
This is indeed the right transformation law for hypermultiplet masses.

It remains to understand how to introduce a nonzero ``global mass'' $\sum_\al m_\al$.
On the ``electric'' side one needs to consider brane configurations in a nontrivial
background geometry~\cite{Witten}. This should correspond to a certain deformation
of Eqs.~(\ref{master'}). In fact, there is one obvious deformation: one may introduce
FI terms for the gauge group ${\cal G}_0$ on the right-hand-side of Eqs.~(\ref{master'}),
deforming them to
\begin{eqnarray}\label{master''}
&&F_{z\ov{z}}-[\Phi_z,\Phi_z^\dagger]=0, \nn \\
&&\ov{D}\Phi_z=-\frac{\pi}{RL}\sum_{\al=1}^k \delta^2 (z-z_\al)\ \diag(m_\al,-M,\ldots,-M).
\end{eqnarray}
Here $M$ is a complex parameter. Note that we did not introduce an FI deformation into the
the first of Eqs.~(\ref{master''}), in agreement with our requirement that the Wilson
lines around the punctures be trivial. Thus, although the FI parameter is three-component
in general, in our case it has only two nonzero components (which we combined into a
complex parameter $M$).
Furthemore, we took all $k$ FI terms to be the same; allowing them to be different
does not give anything new, as one can make a change of variables in Eqs.~(\ref{master''})
which will make
them all equal~\cite{KS}. Therefore we have just one complex deformation parameter $M$.
It is natural to assume that it corresponds to the ``global mass'' on the ``electric'' side. 
It certainly has the right transformation properties with respect to $U(1)_R$.
The identification of $M$ with the ``global mass'' is also in agreement with the general 
correspondence between masses in the ``electric'' theory
and FI terms in the ``magnetic'' theory~\cite{IS,Berkeley,HW}.
Note also that although in $d=3$ the mass parameter has three real components, in a theory
obtained by a straigtforward compactification from $d=4$ one of the components 
(the so-called real mass) is zero. This agrees with the fact that our FI deformation has only
two real components. But the most direct way to see how $M$ is related to the ``global mass''
is to consider the trace part of
Eqs.~(\ref{master''}), and use the fact that the sum of the residues of a meromorphic function
${\rm Tr}\ \Phi$ on a torus must be zero. Then we see that the condition 
$\sum_{\al=1}^k m_\al=0$ must be replaced by $\sum_{\al=1}^k m_\al=k(n-1)M.$

\subsection{The Decompactification Limit}

So far we discussed the Coulomb branch $X_0$ of the $SU(n)^k\times U(1)$ theory compactified
on ${\bf R}^3\times S^1$. We showed that $X_0$ is given by the moduli space of $U(n)$ 
Hitchin equations on a torus with $k$ punctures with residues for $\Phi_z$ of a particular 
kind. We now wish to decompactify $S^1$,
i.e. take $R\ra\infty$. The main idea here is that in the limit $R\ra \infty$ we no longer 
think of $X_0$ as the moduli space of the Coulomb branch endowed with a hyperk\"ahler metric,
but rather as the total space of the Seiberg-Witten fibration (see section 2). 
This amounts to picking
a complex structure on $X_0$ in which it looks like a bundle over ${\bf C}^r$
with fibers being abelian varieties of complex dimension 
$r$.\footnote{In our case $r=kn-k+1$.}
Once the complex structure is fixed, the three K\"ahler forms of $X$, $\omega_1,\omega_2,$ 
and $\omega_3$, can be thought of as a $(1,1)$ K\"ahler form $\omega_r=\omega_1$ and a 
complex symplectic form $\omega=\omega_2+i\omega_3$. Taking $R\ra\infty$ is achieved by 
forgetting $\omega_r$.
Then $X_0$ becomes a complex manifold fibered by $A_r$ over ${\bf C}^r$ and equipped with 
$\omega$. This data provides a solution of the $d=4$ theory in the manner explained in 
section 2. In effect, the solution of the four-dimensional theory is obtained by forgetting
part of the solution of its compactified version.

The distinguished complex structure on $Y$ is easily identified in our case: it is the complex
structure which acts on $z,A_\oz,\Phi_z,$ and all $Q$'s and $\tilde{Q}$'s 
by multiplication by $i$. This follows from considering which $U(1)$ subgroup of
the three-dimensional $SU(2)_R$ symmetry survives the limit $R\ra \infty$.
The corresponding complex structure on $X_0$ can be computed in the following 
(rather standard) manner: using the distingusihed complex structure on $Y$ one can naturally
split the hyperk\"ahler moment
map equations into ``complex'' and ``real'' equations,
discard the real ones, and consider the solutions of the complex equations modulo
the complexified gauge group. The new moduli space is identical to $X_0$ as a complex
manifold. In mathematical terms, one substitutes the holomorphic symplectic quotient for the
hyperk\"ahler quotient. This procedure is quite common in field theory: there one often 
treats an $N=2$ theory as an $N=1$ theory and uses the fact that the classical moduli space
of an $N=1$ theory is the space of solutions of the F-flatness conditions modulo the
complexified gauge group. This allows one to avoid solving $N=1$ D-flatness conditions.
Of course, the remaining F-flatness conditions are nothing but the ``complex'' part of
the $N=2$ D-flatness conditions. In our case the complex equations are
\begin{equation}\label{cmaster}
\ov{D}\Phi_z=-\frac{\pi}{RL}\sum_{\al=1}^k \delta^2 (z-z_\al)\ \diag(m_\al,-M,\ldots,-M).
\end{equation}
As a complex manifold $X_0$ is the space of solutions of this equation modulo the complexified
gauge group $GL(n,{\bf C})$. More precisely, the complexfied gauge group ${\cal G}_0^c$
consists of $GL(n,{\bf C})$ gauge transformations reducing to $GL(n-1,{\bf C})$ at 
$z=z_\al, \al=1,\ldots,k$.

Eq.~(\ref{cmaster}) describes a complex integrable system, just as ordinary Hitchin
equations without punctures.
Indeed, consider the phase space $Y$ which is 
$T^*{\cal M}$, the holomorphic cotangent bundle of the space of $GL(n,{\bf C})$ connections
on $T^2$. This space is parametrized by all pairs $(A_\oz,\Phi_z)$.
$T^*{\cal M}$ has a natural (complex) symplectic
structure, as any cotangent bundle. This makes $Y$ a complex integrable system. 
The set of all $\Phi_z$'s can be thought of as the space of action variables.
Then Eq.~(\ref{cmaster}) says that $X_0$ is a symplectic reduction of $Y$ with respect to 
${\cal G}_0^c$. Hence $X_0$ has a complex symplectic structure and is also integrable. 

If we set $k=1$, we get the system discussed in Ref.~\cite{DW}, with precisely the right 
residue for $\Phi_z$, except that in Ref.~\cite{DW} the gauge group was restricted to be
$SL(n,{\bf C})$.
This slight difference is due to the fact that we are solving a $U(n)$ gauge theory
rather than $SU(n)$. Thus we finally see why this particular Hitchin system gives the 
solution of the $N=2$ $U(n)$ gauge theory with a massive adjoint hypermultiplet.

\subsection{Comparison with M-theory Curves}
In this section we show that for any $k\geq 1$ our solution is equivalent to
that obtained in Ref.~\cite{Witten}. We regard the $T^2$ of the impurity theory
as an elliptic curve $\Sigma$. It was shown in section 3 that its modular parameter is
the microscopic gauge coupling of the ``electric'' theory. Recall that a point in $X_0$ can 
be thought of as a holomorphic $GL(n,{\bf C})$-bundle $E$ over $\Sigma$ together with $\Phi$, 
a section of $K_\Sigma\otimes {\rm End} E$, where $K_\Sigma$ is the canonical bundle of 
$\Sigma$~\cite{DW}. (Simply put, $\Phi_z$ is an adjoint-valued $(1,0)$ form). 
For any point $(E,\Phi)\in X_0$ we consider an $n$-fold cover of $\Sigma$ given by
\begin{equation}
\det(t-\Phi)=0,
\end{equation}
where $t$ takes values in $K_\Sigma$. This gives a Riemann surface $C$. In our case $\Sigma$ 
is an elliptic curve, so $K_\Sigma$ is trivial, and one can think of $t$ and $\Phi$ as
a complex function and an adjoint-valued field, respectively. The coefficients of the 
polynomial $\det(t-\Phi)$ are gauge-invariant polynomials in $\Phi$. By virtue of Hitchin
equations, they are meromorphic differentials on $\Sigma$. They Poisson-commute and 
therefore are the action variables of the integrable system represented by $X_0$. 
Thus we have a projection $X_0\ra {\bf C}^r$, where $r$ is the number of action variables.
It will be shown in end of this section that $r=kn-k+1$.
The angle variables parametrize the fiber of this projection.
As a matter of fact, the fiber is the Jacobian of $C$~\cite{DW}. Indeed, given $(E,\Phi)$ 
and the curve $C$ we have a natural holomorphic line bundle over $C$ whose fiber over 
$(t(z),z), z\in \Sigma$, consists of the eigenvectors of $\Phi(z)$ with eigenvalue $t(z)$.
Conversely, given a line bundle
${\cal L}$ over $C$ we can ``project it down'' to $\Sigma$ and obtain a rank $n$ holomorphic
vector bundle $E$ on $\Sigma$ (in mathematical terminology, $E$ is the direct image sheaf
of ${\cal L}$). Thus a point in $X_0$ can also be thought of as a curve $C$ together with
a point in the Jacobian of $C$. Recalling the discussion of section 2, we conclude that
$C$ is the Seiberg-Witten curve for the gauge system in question. So all we have to do
is to compare $C$ with the curve derived for the same gauge theory in Ref.~\cite{Witten}.
This is very easy to do. The curve $C$ is explicitly given by the equation of degree $n$ 
in $t$:
\begin{equation}\label{eqforC}
t^n-f_1 t^{n-1}+f_2 t^{n-2} -\cdots +(-1)^n f_n =0,
\end{equation}
where $f_i$ are invariant polynomials of $\Phi$. As a consequence of Hitchin equations,
$f_i$ are meromorphic functions on $\Sigma$ with poles at $z=p_\al,\al=1,\ldots,k$. 
Near $z=p_\al$ $\Phi$ has a pole with residue proportional to 
$$\diag(m_\al,-M,\ldots,-M),$$ 
therefore $f_i$ has a pole of degree $i$ there. Let
$r(z)$ be a meromorphic function on $\Sigma$ with a simple pole with residue $iM/(2RL)$ at
each of $p_\al$  and a simple pole with residue $-ikM/(2RL)$ at some other point of $\Sigma$
which we call $p_\infty$. Let us define a new variable $\tilde{t}=t-r(z)$, and rewrite 
Eq.~(\ref{eqforC}) in terms of $\tilde{t}$:
\begin{equation}\label{eqforC'}
\tilde{t}^n-\tilde{f}_1 \tilde{t}^{n-1}+\tilde{f}_2 \tilde{t}^{n-2} -\cdots +(-1)^n 
\tilde{f}_n =0,
\end{equation}
It follows from the form of the residues of $\Phi$ that 
at $z=p_\al$ precisely one root of this equation has a simple pole, hence
the new coefficients $\tilde{f}_i$ have simple poles at $z=p_\al, \al=1,\ldots,k$. Also, 
since $f_i,i=1,\ldots,n$ were nonsingular at $p_\infty$, inevitably $\tilde{f}_i$ has a 
pole of degree $i$ at $p_\infty$. Now in Ref.~\cite{Witten} the curve for the 
$SU(n)^k\times U(1)$ gauge theory was specified by the following two requirements:

(1) It is an $n$-fold cover of $\Sigma$ given by Eq.~(\ref{eqforC'}) where
$\tilde{f}_i$ are meromorphic functions with simple poles at $p_\al,\al=1,\ldots,k$ and
poles of order $i$ at some fixed $p_\infty$.

(2) There is a change of variables $\tilde{t}=t-r(z)$, with $r$ being a meromorphic 
function with a simple pole at $p_\infty$, such that when the curve is expressed in terms
of $t$, as in Eq.~(\ref{eqforC}), the coefficients $f_i$ are nonsingular at $p_\infty$.

We see that our solution agrees with that found in Ref.~\cite{Witten}. 

This description of the complex structure of the moduli space of Hitchin
equations makes it easy to show that $X_0$ has complex dimension $2(kn-k+1)$. It is
sufficient to show that the space of curves $C$ satisfying conditions (1),(2) 
has complex dimension $kn-k+1$, since the total space $X_0$ has dimension twice that.
Since the residues of $r(z)$ are fixed in terms of $M$, $r(z)$ it determined up to an 
additive constant. Furthemore, the coefficient $\tilde{f}_i$ is a meromorphic function
with $k$ simple poles at $z=p_\al$ and a pole of order $i$ at $z=p_\infty$.
The singular part of the Laurent
expansion of $\tilde{f}_i$ at $z=p_\infty$ is determined by $r(z)$, while the residues
at $k$ simple poles are free to vary. Thus $\tilde{f}_i$ depends on $k$ free parameters. 
One exception is $\tilde{f}_1$, since its residues are expressed in terms of mass 
parameters $m_\al$, while its constant part can be removed by a shift of $t$. Thus there is 
a total of $k(n-1)$ parameters in $\tilde{f}_i, i=1,\ldots,n$. Adding a single parameter
in $r(z)$, we get a grand total of $kn-k+1$ parameters, as claimed.
 
\section{Discussion}
In this paper we solved some finite $N=2, d=4$ theories compactified on a circle of radius 
$R$ using a version of mirror symmetry. The Coulomb branch $X_0$ is hyperk\"ahler manifold 
given by the moduli space of solutions of Hitchin equations on a torus with punctures. 
The modular parameter of the torus is the gauge coupling of the four-dimensional theory. 
Remarkably one can determine~\cite{KS} the precise behavior of the Higgs field $\Phi$ at 
the punctures using only the familiar D-brane technology. For a gauge group of rank 
$r$ $X_0$ looks like a $2r$-dimensional torus $A_r$ of size $1/\sqrt{R}$ fibered over 
${\bf R}^{2r}$. The reason that the size of $A_r$ scales as $1/\sqrt{R}$ is due to the 
fact that the size of $T^2$ on which the Hitchin system lives is of order $\sqrt{R}$, 
and the moduli living in $A_r$ essentially come from Wilson lines around $T^2$. In 
the decompactification limit $R\ra\infty$ the torus $A_r$ of the Coulomb 
branch shrinks to zero size, and one is only interested in its complex structure, 
as it gives the Seiberg-Witten solution of the four-dimensional gauge theory. 
The complex structure is easily computed, as it is encoded in the spectral curve of 
the Hitchin system. This spectral curve is precisely the curve describing the geometry
of the M5-brane in the approach of Ref.~\cite{Witten}. Thus we may say that we have a
``matrix'' description of these M5-brane configurations.

In this paper we only discussed elliptic models, but extension to the case of noncompact 
$x^6$ should be straightforward, provided all the beta-functions are zero.
Presumably, one will obtain a Hitchin system on a cylinder with appopriate boundary 
conditions at infinity. Inclusion of orientifold four-planes parallel to D4-branes can  
also be easily accomplished.
A more challenging task is to extend the method to asymptotically free theories.
If the number of D4-branes jumps as one crosses the NS5-branes, the results of Ref.~\cite{KS}
are not directly applicable. Furthemore, although the number of D4-branes jumps, the 
M5-brane is still described by a single curve $C$ which covers the $(x^6,x^{10})$ 
cylinder a fixed number of times which is independent of $x^6$~\cite{Witten}. 
From the ``electric'' point of view, this is explained by the bending of the NS5-branes, 
which provide the missing sheets of the cover. It is not clear how to incorporate this 
effect on the ``magnetic'' side. 

Finally, we would like to comment on the relation of our approach to that of Ref.~\cite{SO}.
On the one hand, in Ref.~\cite{SO} the Coulomb branch of the (compactified) $U(n)^k$ gauge 
theory with matter content described by the $A_{k-1}$ quiver was argued to coincide with the 
moduli space of $n$ $U(k)$ instantons on ${\bf R}^2\times T^2$. On the other hand,
we showed that this same Coulomb branch is given by the moduli space of Eqs.~(\ref{master})
(more precisely, the moduli space of Eqs.~(\ref{master'})).
The two statements are equivalent, because Eqs.~(\ref{master}) are the Nahm transform
of instanton equations on ${\bf R}^2\times T^2$~\cite{KS}. Recall also that in 
Ref.~\cite{Witten} D4-branes suspended between NS5-branes were described macroscopically 
as vortices in the worldvolume theory of NS5-branes. It is hard to make sense of this 
picture on the ``microscopic'' level, as there is no sensible theory of a nonabelian 
two-form potential which supposedly should describe parallel NS5-branes. 
Morally speaking, we showed that after a mirror transform these vortices admit a 
microscopic description as instantons on ${\bf R}^2\times T^2$.

\renewcommand{\thesection}{Acknowledgements}
\section{}

I would like to thank S. Cherkis, A. Hanany, S. Sethi, A. Uranga,
and E. Witten for useful discussions. I am also grateful to E. Witten for reading
a preliminary draft of this paper.


\begin{thebibliography}{99}

\bibitem{Witten} E. Witten, ``Solutions Of Four-Dimensional Field Theories Via M Theory,''
Nucl. Phys. {\bf B500}, 3-42 (1997).

\bibitem{GKMMM} A. Gorskii et al., ``Integrability And Seiberg-Witten Exact Solution,''
Phys. Lett. {\bf B355}, 466 (1995).

\bibitem{MW} E. Martinec and N. Warner, ``Integrable Systems And Supersymmetric 
Gauge Theory,'' Nucl. Phys. {\bf B459}, 97-112 (1996).

\bibitem{DW} R. Donagi and E. Witten, ``Supersymmetric Yang-Mills Theory And Integrable 
Systems,'' Nucl. Phys. {\bf B460}, 299-334 (1996).

\bibitem{Hitchin} N. Hitchin, ``Stable Bundles And Integrable Systems,'' Duke Math. J. 
{\bf 54}, 91 (1987).

\bibitem{Russians} A. Gorskii et al., ``N=2 Supersymmetric QCD And Integrable Spin Chains:
Rational Case $N_f<2N_c$,'' Phys. Lett. {\bf B380}, 75-80 (1996);
A. Gorsky, S. Gukov, and A. Mironov., ``Multiscale N=2 SUSY Field Theories, Integrable
Systems, And Their Stringy/Brane Origin -- I,'' hep-th/9707120. 

\bibitem{DM} E. Markman, ``Spectral Curves And Integrable Systems,'' Comp. Math. {\bf 93}, 
255 (1994); R. Donagi, L. Ein, and R. Lazarsfeld, ``A Non-Linear Deformation Of The Hitchin
Dynamical System,'' alg-geom/9504017.

\bibitem{Gukov} S. Gukov, ``Seiberg-Witten Solution From Matrix Theory,'' hep-th/9709138.

\bibitem{IS} K. Intriligator and N. Seiberg, ``Mirror Symmetry In Three-Dimensional
Gauge Theories,'' Phys. Lett. {\bf B387}, 513-519 (1996). 

\bibitem{SW} N. Seiberg and E. Witten, ``Electric-Magnetic Duality, Monopole
Condensation, And Confinement In $N=2$ Supersymmetric Yang-Mills Theory,''
Nucl. Phys. {\bf B426} 19 (1994);
``Monopoles, Duality, And Chiral Symmetry Breaking In $N=2$ Supersymmetric QCD,''
Nucl. Phys. {\bf B431} 484 (1994).

\bibitem{Berkeley} J. de Boer et al., ``Mirror Symmetry In Three-Dimensional Gauge Theories,
Quivers, And D-Branes,'' Nucl. Phys. {\bf B493}, 101-147 (1997).

\bibitem{HW} A. Hanany and E. Witten, ``Type IIB Superstrings, BPS Monopoles, And
Three-Dimensional Gauge Dynamics,'' Nucl. Phys. {\bf B492}, 152-190 (1997).

\bibitem{KS} A. Kapustin and S. Sethi, ``The Higgs Branch Of Impurity Theories,''
hep-th/9804027.

\bibitem{HKLR} N. J. Hitchin, A. Karlhede, U. Lindstr\"om, and M. Ro\v{c}ek,
``Hyperk\"ahler Metrics and 
Supersymmetry,'' Comm. Math. Phys. {\bf 108} 535-589 (1987).

\bibitem{SO} O. Ganor and S. Sethi, ``New Perspectives On Yang-Mills Theories With
Sixteen Supersymmetries,'' hep-th/9712071. 


\end{thebibliography}
\end{document}